\documentstyle[epsfig]{elsart}

\begin{document}

\runauthor{L. Benet et al.}

\begin{frontmatter}
\title{Effect of a finite--time resolution on Schr\"{o}dinger 
cat states in complex collisions}

\author[ccf,cic]{L. Benet},
\author[ccf,cnp,dtp]{S.Yu. Kun},
\author[imp]{Wang Qi},
\author[inr]{V. Denisov}

\address[ccf]{Centro de Ciencias F\'{\i}sicas, National University of 
  Mexico (UNAM), Cuernavaca, Mexico}
\address[cic]{Centro Internacional de Ciencias, Cuernavaca, Mexico}
\address[cnp]{Center for Nonlinear Physics, RSPhysSE, The Australian 
  National University, Canberra ACT 0200, Australia}
\address[dtp]{Department of Theoretical Physics, RSPhysSE, The Australian 
  National University, Canberra ACT 0200, Australia}
\address[imp]{Institute of Modern Physics, Chinese Academy of Sciences,
  Lanzhou 730000, China}
\address[inr]{Institute for Nuclear Research, 03680 Kiev, Ukraine}

\date{\today}

\begin{abstract}
We study the effect of finite time resolution on coherent superpositions 
of rotating clockwise and anticlockwise wave packets in the regime of 
strongly overlapping resonances of the intermediate complex. Such highly 
excited deformed complexes may be created in binary collisions of 
heavy-ions, molecules and atomic clusters. It is shown that time
averaging reduces the interference fringes acting effectively 
as dephasing. We propose a simple estimate of the ``critical'' time 
averaging interval. For the time averaging intervals bigger than the 
critical one the interference fringes wash out appreciably. This is 
confirmed numerically. We evaluate minimal energy intervals for measurements 
of the excitation functions needed to observe the Schr\"{o}dinger Cat
States. These should be easily observable in heavy-ion
scattering. Such an experiment is suggested. 
\end{abstract}

\end{frontmatter}

%----------------------------------------------------------------------

Time dependence of the decay of an intermediate complex provides
the most detailed information on the collisional molecular dynamics.
In order to get such an information experimentally one needs to
arrange for a ``clock''. The clock should be switched on at the
moment when the collision partners start to interact. It should also 
fix the moment of the decay of the intermediate complex. For unimolecular 
reactions this problem was successfully solved using pump and probe 
laser pulses~\cite{Zewail95,Zewail99}. The pump pulse swithches on the 
clock while the probe pulse determines if the intermediate complex has 
decayed or not. By changing continuously the time delay between the pumpe 
and probe laser pulses one measures the time dependence of the decay of 
the intermediate complex.

Unfortunately such a real--time monitoring of collisional dynamics
can not be applied for bimolecular collisions unless these are initiated
by a laser pulse~\cite{Zewail95,Zewail99}. However it was 
shown~\cite{Kun01,Kun02} that measurements of the cross sections with 
pure energy resolution enable one to extract the information on the time 
and scattering angle intensity of the decay of the intermediate complex.
This information is equivalent to that obtained in real--time
pumpe/probe laser pulses experiments provided the relative contribution
of the direct (fast) processes in the energy averaged cross section
is appreciable. Then the energy dependence of the cross section is mainly 
determined by interference between energy smooth direct reaction amplitude 
and the fluctuating one corresponding to time delayed processes.

The possibility of formation of coherent superpositions of wave packets 
rotating clockwise and anticlockwise in the regime of strongly overlapping 
resonances of the intermediate complex has recently been 
suggested~\cite{Kun03}. In this regime the rotational period is much 
shorter than the Heisenberg time of the quasibound complex. The wave 
packets are stable and do not spread appreciably due to the small width 
of spin off--diagonal phase relaxation as compared to the inverse
rotational period and inverse average life--time of the intermediate
complex. Such highly excited complexes may be created in binary collisions 
of heavy--ions, molecules and atomic clusters provided the double--folding 
potential between the collision partners has a pocket. The calculations 
support the existence of such pockets for heavy--ion~\cite{Greiner95} 
and atomic cluster~\cite{Engel93,Schmitt94} systems. When the relative 
kinetic energy of the colliding partners dissipates into intrinsic 
excitation they drop into this pocket, forming a highly excited 
intermediate complex.

It has been unambiguously demonstrated in double slit like 
experiments~\cite{Arndt99} that fullerene beams do indeed represent
waves. This supports the conceptual similarity in the quantum mechanical 
treatment of atomic clusters, i.e. mesoscopic, collisions and heavy--ion 
collisions~\cite{Schmidt92}.

It should be noted that stable wave packets in highly--excited
many--body systems have been identified and associated with 
the existence of invariant manifolds in the classical phase space of 
the system~\cite{Papen98,Papen01}.

It can be seen from Fig.~1 of Ref.~\cite{Kun03} that the angular 
period of the interference fringes is about $\pi/I$, where $I$ is the 
average spin of the intermediate complex. In order to detect the effect 
experimentally, one must measure the cross section with angular resolution 
better than $\pi/I$. Panels~(f) and~(g) in Fig.~1 of Ref.~\cite{Kun03}
also provide an indication of a strong time dependence of the interference 
fringes. Therefore an unaswered question is of what time resolution $\Delta t$ 
is necessary to detect the effect. This question is of a crutial importance
for planning the experiment, since this required time resolution
determines a minimal energy interval $\Delta E=\hbar/\Delta t$ 
on which the cross section measurements must be carried out in order to 
resolve the interference fringes.

First we suggest a simple estimate for the time averaging interval
$\Delta t_{cr}$ obtained from the required angular resolution, $\Delta \theta
\simeq \pi/I$, needed to resolve angular dependence of the interference 
fringes. The time it takes the intermediate complex to rotate by the
angle $\pi/I$ is $\pi/ I\omega=T/2I$, where $T=2\pi/\omega$ is the
period of rotation and $\omega$ is the angular velocity of the intermediate
system. This suggests $\Delta t_{cr}\simeq T/2I$. In what follows this 
simple estimate will be tested and confirmed numerically.

Following Ref.~\cite{Kun03} we consider spinless collision partners 
in the entrance and exit channels. The time and angle dependent intensity
of the decay, $P(t,\theta )$, has been obtained~\cite{Kun03} by 
summing over very large number of strongly overlapping resonance levels,
 ${\bar t}\ll\hbar /D$, where ${\bar t}$ is the average life time and 
$D$ is the average level spacing of the intermediate complex. As a 
result $P(t,\theta )$ has the form:
\begin{eqnarray}
P(t,\theta )& \propto & H(t)\exp(-t/{\bar t})\sum_{JJ^\prime }
(2J+1)(2J^\prime +1)[W(J) W(J^\prime )]^{1/2} \nonumber \\
 & & \exp[i(\Phi -\omega t) (J-J^\prime ) -
\beta |J-J^\prime |t/\hbar ] P_J(\theta )P_{J^\prime}(\theta ).
\label{eq1}
\end{eqnarray}
Here $H(t)$ is the Heaviside step function, $\beta$ is the spin relaxation
width, $\omega$ is the angular velocity of the coherent rotation,
$\Phi$ is the deflection given by the total spin ($J$) derivative
of the potential (direct reaction) phase shifts, and the $P_J(\theta)$ 
are Legendre polynomials. The partial average reaction probability is 
taken in the $J$--window form, 
$W(J)= <|\delta S^J(E)|^2>\propto\exp[-(J-I)^2/d^2]$,
where $I$ is the average spin and $d\ll I$ is the $J$--window width. 
This means that we assume a peripheral character of the collision. 

In order to study the effect of the time averaging on the interference
fringes we average $P(t,\theta)$ in~(\ref{eq1}), with a constant weight, 
on the time interval from $t-\Delta t/2$ to $t+\Delta t/2$. This 
integration is elementary. Then we calculate the time averaged $P(t,\theta)$ 
numerically.

Like in Ref.~\cite{Kun03}, we calculate $P(t,\theta )$ with the set of 
parameters obtained from the description~\cite{Kun01} of the 
experimental cross section energy autocorrelation 
functions~\cite{Ghosh87} for $^{12}$C+$^{24}$Mg elastic and inelastic 
scattering at $\theta=\pi$~\cite{Mermaz81}. For these collisions the 
analysis~\cite{Kun01} indicates a formation of a stable rotational 
wave packets in spite of strong overlap of resonance levels in the 
highly--excited intermediate molecule. The set of parameters 
is~\cite{Kun01}: $\Phi=0$, $d=3$, $I=14$, $\beta=0.01$ MeV, 
$\hbar\omega =$1.35 MeV, and ${\bar t}=2.2\times 10^{-21}$ sec. It should 
be noted that analysis of the angular distributions~\cite{Mermaz81}
supports a peripheral reaction mechanism, i.e. $d\ll I$, in the 
$^{12}$C+$^{24}$Mg elastic scattering.

\begin{figure}
\noindent\centerline{
\psfig{figure=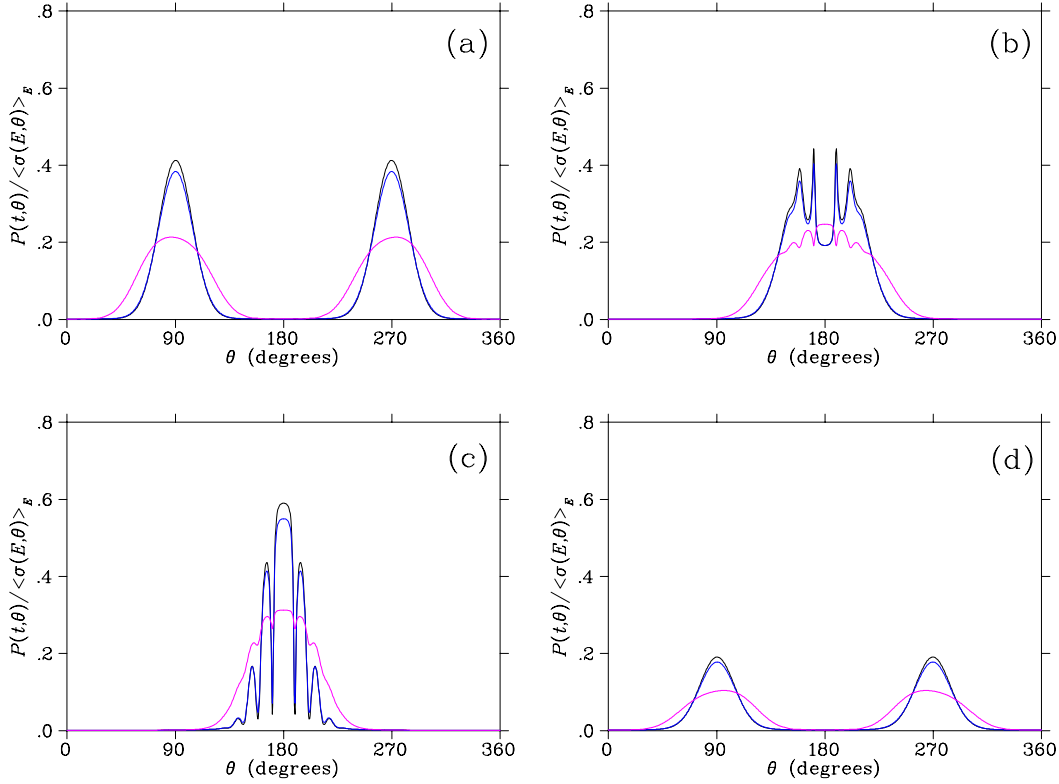,height=\textwidth,angle=90}}
\caption{
\label{fig1}
(Color online) Time and angular dependence of decay intensity  
 of the highly--excited intermediate complex for different time 
averaging intervals: (a)~$t=T/4$; (b)~$t=7T/16$; (c)~$t=T/2$;
(d)~$t=3T/4$. $T$ is the period of one complete revolution of the complex.
Black lines correspond to ideal time resolution $\Delta t=0$;
blue $\Delta t=\Delta t_{cr}$; magenta $\Delta t=5\Delta t_{cr}$. The 
bigger $\Delta t$ the more the spread of the wave packets and the 
stronger reduction of the interference fringes.}
\vskip0.5cm
\end{figure}

In Fig.~\ref{fig1} we demonstrate the dependence of the quantity
$AP(t,\theta )/<\sigma(E,\theta)>$ on the time averaging interval for 
the four moments of time. Here 
$<\sigma(E,\theta)>\propto \int_0^\infty dtP(t,\theta )$ is the
energy average differential cross section for the time-delayed 
collision. In Fig.~\ref{fig1}, $P(t,\theta )$ is scaled with 
$<\sigma(E,\theta)>$ for the reason discussed in Ref.~\cite{Kun03}. The 
constant $A$ is derived from the condition 
$AP(t=0,\theta=0 )/<\sigma(E,\theta=0)>=1$, where 
$P(t=0,\theta=0 )$ is not averaged over the time.

% Figure 1

From Fig.~\ref{fig1} one can see that an increase of the time averaging 
interval results in a reduction of the amplitude of the interference 
fringes.

% Figure 2
\begin{figure}
\noindent\centerline{
\psfig{figure=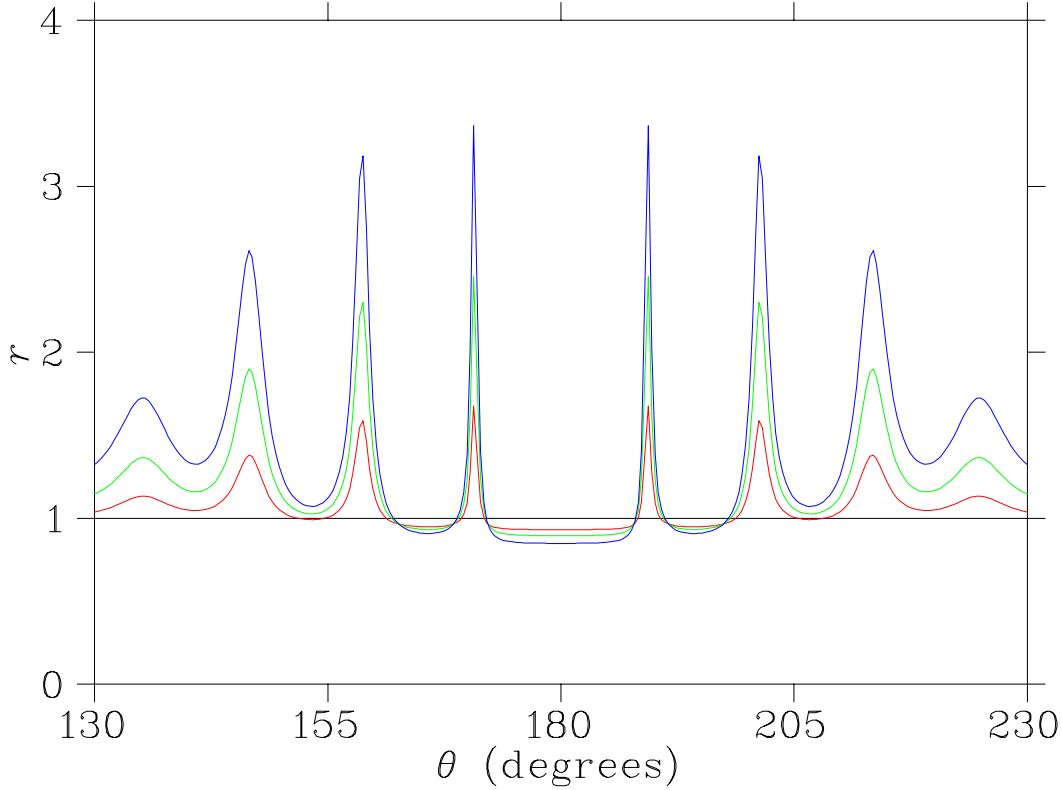,height=\textwidth,angle=90}}
\caption{
\label{fig2}
(Color online) Ratio 
$r=P^{\Delta t}(t=T/2,\theta )/ P^{\Delta t=0}(t=T/2,\theta )$
of the time averaged  $P^{\Delta t}(t=T/2,\theta )$
and the $P^{\Delta t=0}(t=T/2,\theta )$ corresponding to ideal time 
resolution. The time averaging has been performed around $t=T/2$
with $T$ being the period of one complete revolution of the complex.
Black color is for ideal time resolution $\Delta t=0$; red  
$\Delta t=t_{cr}$; green $\Delta t=3t_{cr}/2$; blue $\Delta t=2t_{cr}$. 
The minima of the curves reflect suppression of the peaks of the 
interference fringes (see Fig.~\ref{fig1}) due to the time
averaging. The maxima signify filling in of the interference
deeps (see Fig.~\ref{fig1}) due to the time averaging. The bigger the 
time averaging interval $\Delta t$ the greater deviation from the 
constant unity.}
\vskip0.5cm
\end{figure}

In Fig.~\ref{fig2} we plot a ratio, 
$r=P^{\Delta t}(t=T/2,\theta )/ P^{\Delta t=0}(t=T/2,\theta )$,
of the time averaged  $P^{\Delta t}(t=T/2,\theta )$
and the $P^{\Delta t=0}(t=T/2,\theta )$ corresponding to ideal time 
resolution. The time averaging has been performed around $t=T/2$.
At this moment, corresponding to a half of a rotational period, 
the two wave packets overlap around $\theta=\pi$~\cite{Kun03}. The 
constant unity corresponds to the ideal time resolution. The minima in 
Fig.~\ref{fig2} reflect  the suppression of the peaks of the inteference 
fringes due to the time averaging. One can see that this suppression is 
not appreciable. The maxima in Fig.~\ref{fig2} signify filling in of 
the interference minima due to the time averaging. One can see that, for 
$\Delta t=\Delta t_{cr}= T/2I$, the interference deeps are reduced
by a factor $\leq$1.5 only. However, for  
$\Delta t=1.5\times\Delta t_{cr}$, this reduction is given by 
a factor of about 2.5, and, for $\Delta t=2\times\Delta t_{cr}$,
by a factor of about 3.5. Therefore, for the time averaging intervals 
bigger than the critical one, the interference fringes wash out fast 
and appreciably. 

Thus, for the critical time averaging interval, the reduction of the 
interference maximum/minimum ratio is given by a factor of about 1.5. Since, 
for the ideal time resolution, the interference maximum/minimum ratio 
exceeds one order of magnitude (see Fig.~\ref{fig1}), this ratio is 
still of about of factor 6 after the averaging over $\Delta t_{cr}$. In 
order to have the time resolution $\Delta t_{cr}$ one has to measure the 
cross section on the energy interval 
$\Delta E\simeq \hbar/\Delta t_{cr}=(I/\pi )\hbar \omega$. For the above 
considered $^{12}$C+$^{24}$Mg elastic and inelastic 
scattering~\cite{Ghosh87,Mermaz81} we have $I=14$ and 
$\hbar\omega =$1.35 MeV~\cite{Kun01}. This yields 
$\Delta E \simeq 6$ MeV. The measurements of the $^{12}$C+$^{24}$Mg 
elastic and inelastic scattering at $\theta=\pi$~\cite{Mermaz81} and 
the consequent extraction of the cross section energy autocorrelation 
functions~\cite{Ghosh87} were performed on the energy intervals 
bigger than 10 MeV and 9 MeV, respectively. Clearly, the measurements of 
the excitation functions on such energy intervals for many backward 
angles $130^\circ\leq \theta \leq 180^\circ$ with a fine angular
resolution $\Delta\theta\leq \pi/3I\simeq 4-5$ degrees and a fine
angular step would allow to search for Schr\"odinger Cat States in 
$^{12}$C+$^{24}$Mg elastic scattering. Such an experiment would be 
desirable since the Schr\"odinger Cat States predicted~\cite{Kun03}
for $^{12}$C+$^{24}$Mg scattering involve $\sim 10^4$ many--body 
configurations of the highly excited intermediate complex. To the best 
of our knowledge, the internal interactive complexity of these quantum 
macroscopic superpositions dramatically exceeds~\cite{Kun03} all of 
those previously experimentally realized.

\ack
We are grateful to Thomas H. Seligman for useful discussions and 
suggestions. One of us (LB) acknowledges finantial support from the 
projects IN--101603 (DGAPA--UNAM) and 43375-E (CONACyT).

%----------------------------------------------------------------------


\begin{thebibliography}{99}

\bibitem{Zewail95} A.H. Zewail, {\sl Femtochemistry} (World Scientific, 
 Singapore, 1995), Vols. 1 and 2.

\bibitem{Zewail99} A.H. Zewail, Nobel Lecture 1999, in {\sl Nobel Lectures, 
  Chemistry 1996-2000}, editor Ingmar Grenthe (World Scientific, Singapore, 
  2003).

\bibitem{Kun01} S.Yu. Kun, A.V. Vagov, and W. Greiner, Phys. Rev. C {\bf 63}, 
014608 (2001).

\bibitem{Kun02} S.Yu. Kun, A.V. Vagov, L.T. Chadderton, and W. Greiner,  
Int. J. Mod. Phys. E {\bf 11}, 273 (2002).

\bibitem{Kun03} S.Yu. Kun, L. Benet, L.T. Chadderton, W. Greiner, and 
F. Haas, Phys. Rev. C {\bf 67}, 011604(R) (2003).

\bibitem{Greiner95} W. Greiner, J.Y. Park, and W. Scheid, {\sl Nuclear 
Molecules} (World Scientific, Singapore, 1995).

\bibitem{Engel93} E. Engel {\sl et al.}, Phys. Rev. B {\bf 48}, 1862 (1993).

\bibitem{Schmitt94} U.R. Schmitt, E. Engel, and R.M. Dreizler, Phys. Rev. B 
{\bf 50}, 14674 (1994).

\bibitem{Arndt99} M. Arndt, O. Nairz, J. Vos-Andreae, C. Keller, G. van der 
Zouw, and A. Zeilinger,  Nature {\bf 401}, 680 (1999).

\bibitem{Schmidt92} {\sl Nuclear Physics Concepts in the Study of Atomic 
Cluster Physics}, edited by R. Schmidt, H.O. Lutz, and R. Dreizler,
Lecture Notes in Physics Vol. 404 (Springer-Verlag, Heidelberg, 1992). 

\bibitem{Papen98} T. Papenbrock, T.H. Seligman, and H.A. Weidenm\"uller,
Phys. Rev. Lett. {\bf 80}, 3057 (1998)

\bibitem{Papen01} T. Papenbrock and T.H. Seligman, J. Phys. A: Math. Gen. 
{\sl 34}, 7423 (2001).

\bibitem{Ghosh87} B. Ghosh and R. Singh, Pramana J. Phys. {\bf 29}, 1555 
(1987).

\bibitem{Mermaz81} M.C. Mermaz {\sl et al.}, Phys. Rev. C {\bf 24}, 1512 
(1981).


\end{thebibliography}
\end{document}